\documentclass[aps,prd,notitlepage,reprint,onecolumn,showpacs,superscriptaddress,nofootinbib,floatfix]{revtex4-1}

\usepackage{amsmath,amssymb,amsfonts}
\usepackage{graphicx}
\usepackage{verbatim}
\usepackage{pifont}

\usepackage{hyperref}
\usepackage{slashed}
\usepackage{verbatim}
\usepackage{rotate}

\usepackage[normalem]{ulem}
\usepackage{cancel}

\usepackage{graphicx,color}

\newcommand{\be}{\begin{equation}}
\newcommand{\ee}{\end{equation}}
\newcommand{\bi}{\begin{itemize}}
\newcommand{\ei}{\end{itemize}}
\newcommand{\bea}{\begin{eqnarray}}
\newcommand{\eea}{\end{eqnarray}}
\newcommand{\ud}{\mathrm{d}}

\newcommand{\LCperp}{{\scriptscriptstyle \perp}}

\usepackage[T1]{fontenc} \usepackage[latin1]{inputenc}

\begin{document} 

\title{Very Special Relativity as a background field theory}

\author{Anton Ilderton}
\affiliation{Department of Physics, Chalmers University, SE-41296 Gothenburg, Sweden}
\email{anton.ilderton@chalmers.se}

\begin{abstract}
We consider violation of Lorentz invariance in QED induced by a very high frequency background wave. An effective theory is obtained by averaging observables over the rapid field oscillations. This preserves Ward identities and restores translation invariance below the high frequency scale, but only partial Lorentz invariance: we show that the effective theory is C-invariant  SIM(2)--QED in Very Special Relativity. Averaging generates the nonlocal terms familiar from SIM(2) theories, while the short-distance behaviour of the background field fermion propagator generates the infinite number of higher-order vertices of SIM(2)-QED.
\end{abstract}

\maketitle

\section{Introduction}\label{intro}
Effective descriptions of physics beyond the Standard Model can include Lorentz-invariance-violating effects due to the spontaneous breaking of Lorentz symmetry in more fundamental theories~\cite{Colladay1,Colladay2}. Alternatively Lorentz invariance may itself be an effective description of a more fundamental spacetime symmetry: here we will see an example of how these ideas may be related. For a recent review of tests of Lorentz invariance see~\cite{Liberati:review}.

In Very Special Relativity (VSR), the spacetime symmetry group of nature is taken to consist of translations and only a subgroup of the Lorentz group~\cite{original,Cohen2}. The largest such subgroup is the four-dimensional SIM(2), which leaves invariant a null direction $n_\mu$ ($n^2=0$)~\cite{Gibbons}: if we take $n_\mu=(1,0,0,1)$ then SIM(2) is generated by two transverse boosts ($M^{0\LCperp}=\{K^1+J^2,K^2-J^1\}$),  rotations about $z$ ($J^3$) and boosts in the $z$-direction ($K^3$). One of the characteristic features of SIM(2) invariant theories is the appearance of nonlocal terms $n_\mu/n.\partial$. It has been speculated that such terms arise from an unknown medium or \ae ther~\cite{Gibbons,Cheon}: here we show that a toy model for this medium is a very high frequency electromagnetic wave, at least in the context of QED. Specifically we will show that SIM(2)--QED is an effective theory of QED in a high frequency background wave.

This paper is organised as follows. We begin by showing in Sect.~\ref{SEKT:DIRAC} how the SIM(2)--Dirac equation arises as an effective description of classical fermion dynamics in a very high frequency background plane wave. This is established by considering observables averaged over many periods of the rapidly oscillating wave: averaging, which is a nonlocal procedure, generates the nonlocal terms typical of SIM(2) theories. In Sect.~\ref{SEKT:QED} we apply the same approach to establish an effective theory of QED in the same background, showing that averaging turns correlation functions and scattering amplitudes in QED into those of SIM(2)--QED. Ward identities are verified, and the presence of higher-order vertices in SIM(2)--QED is explained in terms of the short-distance behaviour of the background-field fermion propagator. We conclude in Sect.~\ref{SEKT:CONCS} and discuss possible extensions.

\section{Dirac equation}\label{SEKT:DIRAC}
\begin{figure}[t!]
\centering\includegraphics[width=0.4\textwidth]{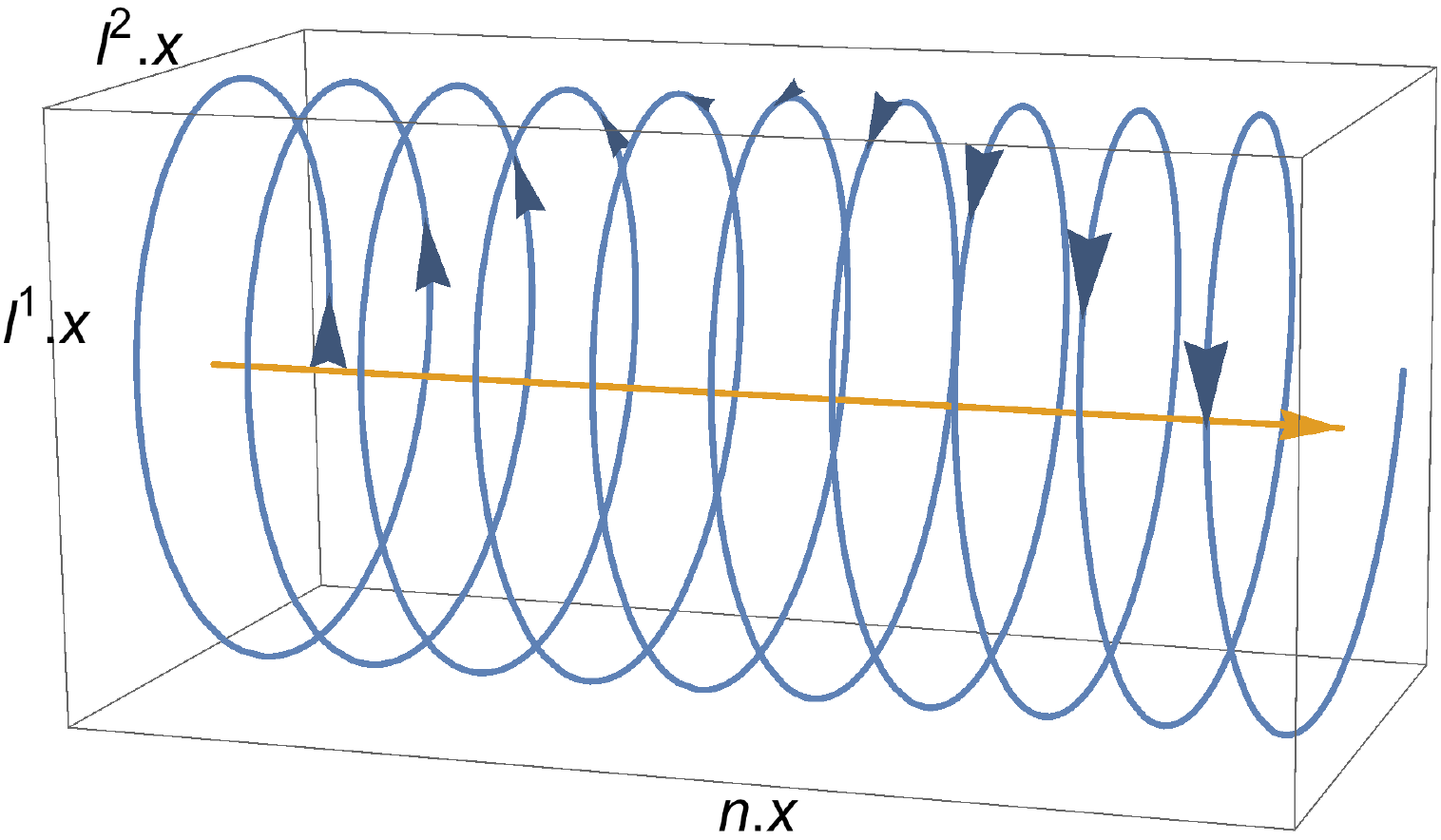}
\caption{\label{FIG:SNITT} Illustration of the exact and effective classical orbits of a particle in the background~(\ref{bakgrund}), as a function of elapsed lightfront time $n.x$. The exact spiral orbit is given by integrating $\pi_\mu$, while the average, linear, orbit is given by integrating $q_\mu$. }
\end{figure}

Consider a background plane wave $F^{\mu\nu}(n.x) = f_j(n.x) \big( n l^j - l^j n\big)^{\mu\nu}$, where $n_\mu$ is lightlike as above, so $n^2=0$, and the two $l^j_\mu$ are spacelike, orthogonal and transverse to $n_\mu$, so $l^j.l^k=-\delta^{ij}$ and $n.l^j=0$. This background is well-studied in the context of intense laser-matter interactions, where it provides a first model of the laser fields; see~\cite{Marklund:2006my,Heinzl:ELI,EKK,DiPiazza:review} for reviews. Here though we imagine $F_{\mu\nu}$ as a `fundamental' (spacetime-dependent) VEV for the gauge field, or as the presence of some fixed coherent state of photons, which introduces a preferred spacetime direction $n^\mu$ but which is homogeneous in the remaining three directions. The classical dynamics of a particle, charge $e$ and mass $m$, in this background are compactly phrased in terms of an effective work done by the field, $a_\mu$, where
\be\label{A-DEF}
	a_\mu(n.x) := e\int\limits_{-\infty}^{n.x}\!\ud\varphi\; f_j(\varphi) l^j_\mu \;.
\ee
Using $a_\mu$, the time-dependent kinetic momentum $\pi_\mu$ of the particle may be parameterised using `lightfront time' $n.x$ as
\be\label{PI}
	\pi_\mu(n.x) = p_\mu  - a_\mu(n.x) + \frac{2p.a(n.x)-a^2(n.x)}{2n.p}n_\mu \;,
\ee
in which the particle has momentum $p_\mu$ before entering the wave, and $p^2 =m^2 = \pi^2$. (Note that $n.p>0$ for massive on-shell particles, so that classical motion is not singular. See also Section~\ref{SECT:LF}.) Imagine now that the background has (in a typical terrestrial lab frame), a very high frequency such that any typical physical process will occur over very many periods of the wave. The effective physical observables will then be averages over these many, rapid oscillations. We illustrate using the momentum (\ref{PI}) and the simplest case of a monochromatic and circularly polarised wave. In this case we can always choose a frame such that $n_\mu=(1,0,0,1)$, as in the introduction, which fixes $l^1_\mu=\delta_\mu^1$ and $l^2_\mu=\delta_\mu^2$. The work done (\ref{A-DEF}) then takes the explicit form, for our chosen field,
\be\label{bakgrund}
	a_\mu(n.x) = \delta m \big(0,\cos (\omega n.x), \sin (\omega n.x), 0\big) \;,
\ee
where the amplitude $\delta m^2 = -a^2$ is positive and constant, and $\omega$ is the high frequency. The momentum averaged over $N$ cycles, $\langle \pi_\mu \rangle$, is independent of $n.x$ and $N$, being equal to
\be\label{q}
	q_\mu := \langle \pi_\mu \rangle  = p_\mu + \frac{\delta m^2 }{2n.p}n_\mu \;.
\ee
The final term in $q_\mu$ is typical of SIM(2)~\cite{Dunn,Cohen2}. The interpretation of (\ref{PI}) and (\ref{q}) is that if we do not probe scales as high as $\omega$ then we do not resolve the field oscillations or their effects, and hence see the particle as having a momentum $q_\mu$ which is on mass-shell at
\be\label{massskiften}
	q^2 = m^2 + \delta m^2 \;,
\ee
whereas if we could resolve the high frequency scale we would see that the particle actually has mass $m$, see Fig.~\ref{FIG:SNITT}. We will confirm below that $m^2+\delta m^2$ is the observed rest mass squared in SIM(2). Note that the wave frequency $\omega$ has dropped out of~$q_\mu$: as in VSR, the direction of the `\ae  theral motion' is $n_\mu$ but we have no access to the velocity/frequency of this \ae ther~\cite{Gibbons}.

To make the connection with SIM(2) explicit consider the Dirac equation. We can take $eA_\mu = a_\mu$ as a gauge potential for the wave (which we note for later is in lightfront gauge $n.A=0$). With this, the solution to the Dirac equation in a plane wave background is~\cite{Volkov}
\be\label{V-SOL-SPINOR}
	\psi(x) = \int\!\ud p\;  \bigg(1 + \frac{ \slashed{n} \slashed{a}(n.x)}{2n.p} \bigg)u_p b_p\phi_p(x) + \bigg(1 - \frac{ \slashed{n} \slashed{a}(n.x)}{2n.p}\bigg)v_p d^\dagger_p \phi_p^\dagger(x)\big|_{a\to -a} \;,
\ee
where $\ud p$ is the on-shell measure, the first/second term in the integrand describes electrons/positrons, the scalar functions $\phi_p$ solve the Klein-Gordon equation,
\be\label{V-SOL-SKALAR}
	\phi_p(x) = \exp \bigg(-ip.x -\frac{i}{2n.p} \int\limits^{n.x}_{-\infty} 2p.a -a^2\bigg) \,,
\ee
with $p^2=m^2$, and the time-dependent spinor
\be\label{U-DEF}
	 u_{\pi(n.x)} :=\bigg(1 + \frac{ \slashed{n} \slashed{a}(n.x)}{2n.p} \bigg) u_p \;,
\ee
obeys a time-dependent version of the usual Dirac condition, $\big(\slashed{\pi}(n.x) - m\big) u_{\pi(n.x)} = 0$. The Volkov solutions (\ref{V-SOL-SPINOR}) become the one-particle wavefunctions in the quantum theory and, writing $\psi_p = u_\pi \phi_p$, recover the classical current via   $\bar{\psi}_{p} \gamma^\mu \psi_p  = \pi^\mu$. The average current is of course~(\ref{q}): what `averaged field' yields this current, and how is it is related to (\ref{V-SOL-SPINOR})? Consider that averaging typically removes rapidly oscillating terms. In the momentum~(\ref{PI}) and current the terms linear in $a_\mu$ are rapidly oscillating and vanish upon averaging, leading to (\ref{q}), while quadratic terms, $a^2(n.x) = -\delta m^2$, are constant (i.e.~slowly varying) and survive averaging. We therefore define an averaged field $\psi^\text{av}$ by dropping the linear terms in (\ref{V-SOL-SPINOR}); this amounts to replacing $\phi_p(x) \to \exp\big(-i q.x \big)$ and, from (\ref{U-DEF}), $u_\pi \to u_p$:
\be\label{psi-av}
	\psi^\text{av}(x) := \int\!\ud p\; b_p u_p e^{-iq.x} + d^\dagger_p v_p e^{iq.x} \;.
\ee
The mode functions of this field, $\psi_p^\text{av}:=u_pe^{-iq.x}$ generate the averaged current as wanted:
\be
	\bar\psi^\text{av}_{p} \gamma^\mu \psi^\text{av}_p = q^\mu \;.
\ee
Applying the Dirac operator, we find that the averaged field obeys the equation of motion
\be\label{EOM}
	\bigg( i\slashed{\partial} - m - \frac{\delta m^2}{2in.\partial}\slashed{n} \bigg) \psi^\text{av}(x) = 0 \;,
\ee
which is the free spin--$\tfrac{1}{2}$ equation of motion in SIM(2) VSR~\cite{Cohen2,Dunn,Lee:2015tcc}. Thus fermion dynamics below the high-frequency scale is effectively described by the SIM(2) Dirac equation. Note that squaring (\ref{EOM}) yields $(\partial^2 +m^2+\delta m^2)\psi^\text{av}=0$, so that the effective mass is $m^2+\delta m^2$, in agreement with (\ref{massskiften}).

\section{Background fields $\to$ SIM(2)-QED}\label{SEKT:QED}
We now turn to the quantum theory and, following the above argument of dropping rapidly oscillating terms, ask what is the gauge-invariant and consistent truncation of QED in a background field which yields $\psi^\text{av}$ as the effective physical degrees of freedom?

The observables of interest are probabilities calculated from $S$-matrix elements, themselves built from correlation functions by amputation where external legs become the one-particle wavefunctions~(\ref{V-SOL-SPINOR}). We start with the fermion propagator in a plane wave~\cite{Volkov,Mitter:Schladming},
\bea\label{GVOLK}
	S(x,y) = i \int\! \frac{\ud^4 p}{(2\pi)^4}  \phi_p(x)\phi^\dagger_p(y)
	\bigg(1+\frac{\slashed{n}\slashed{a}(n.x)}{2n.p}\bigg)\frac{\slashed{p}+m}{p^2-m^2+i\epsilon}\bigg(1+\frac{\slashed{a}(n.y)\slashed{n}}{2n.p}\bigg) \;,
\eea
and make the same averaging replacements as above, dropping terms linear in $a_\mu$ from the exponent and the spinors. It is easily checked that this turns $S$ into the propagator of the averaged spinor functions $\psi^\text{av}_p$. Changing integration variable $p_\mu \to q_\mu$ then shows that the propagator reduces to
\be\label{G2VSR}
	S_\text{vsr}(x-y) = i \int\! \frac{\ud^4 q}{(2\pi)^4} \frac{\slashed{q}+m - \frac{\delta m^2\slashed{n}}{2n.q}}{q^2-m^2-\delta m^2+i\epsilon} e^{-i q.(x-y)} \;,
\ee
which is the spin--$\tfrac{1}{2}$ propagator of SIM(2). Note that although we began with a non-constant background field in (\ref{GVOLK}), translation invariance is restored in (\ref{G2VSR}) below the high-frequency scale of the background (compare also  (\ref{PI}) and (\ref{q})). Hence momentum becomes a good quantum number after averaging, just as it is in VSR~\cite{original}. The propagator poles lie at $m^2+\delta m^2$, which is the observable particle mass in SIM(2).

While the above suggests a close connection between QED in our background and SIM(2)--VSR, we have really only considered the Dirac equation so far. Addressing interactions requires more care, for three reasons. First, the propagator (\ref{GVOLK}) and wavefunctions (\ref{V-SOL-SPINOR}) are gauge-dependent objects, so we must ensure that we do not violate Ward-Takahashi identities as we average out the high-frequency scale. Second, in a general correlation function two fermion propagators will meet at each vertex, which may generate slowly varying terms (quadratic in $a_\mu$ at the same spacetime point) from the $\slashed{a}$ terms which we have dropped in going from (\ref{GVOLK}) to (\ref{G2VSR}). Third, we note that the spin factor in~(\ref{GVOLK}) contains a term~$\sim a(n.x).a(n.y)$. This is rapidly oscillating in general  as the fields are evaluated at different spacetime points, hence we have dropped it when going to (\ref{G2VSR}). However, when $n.x\simeq n.y$ a short-distance expansion of the propagator may include slowly varying terms which are not explicit in (\ref{GVOLK}). We will see that all three of these points are related, and that all need to be accounted for in order to preserve gauge invariance and to show that the consistent effective theory is indeed SIM(2)--QED. Note that because averaging removes terms linear in the charge, $\langle a \rangle \to 0$, while terms quadratic in the charge survive, $\langle a^2 \rangle \not\to 0$,
the SIM(2) theory we obtain will be $\mathcal{C}$-invariant~\cite{Dunn}.
 
\begin{figure}[t!]
	\centering\includegraphics[width=0.6\columnwidth]{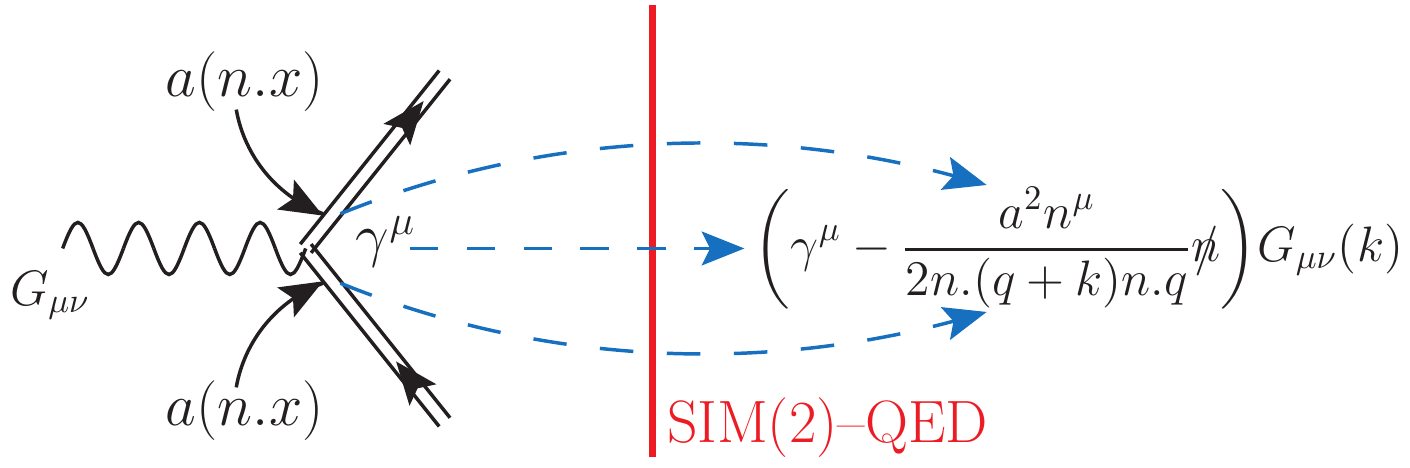}
	\caption{\label{FIG:WARD} Products of background field propagators (double lines, left) across vertices in QED generate slowly varying terms necessary to preserve the Ward-Takahashi identity, and generate the three-point SIM(2)-vertex (right).}
\end{figure}

Consider then a general correlation function, and a vertex within it at spacetime position $x^\mu$. Let the fermion propagators entering/exiting the vertex carry momentum (by which we refer to the integration variable in (\ref{GVOLK})) $p_\text{in}$/$p_\text{out}$ respectively. The spinor structure at this vertex is
\be
	\int\ldots \bigg(1+\frac{\slashed{a}(n.x)\slashed{n}}{2n.p_\text{out}}\bigg) \gamma^\mu  \bigg(1+\frac{\slashed{n}\slashed{a}(n.x)}{2n.p_\text{in}}\bigg) G_{\mu\nu}\ldots 
\ee
where $G_{\mu\nu}$ is the photon propagator in a general covariant gauge. There are two slowly-varying terms here: the $\gamma^\mu$ term, and a term quadratic in the external field arising from contractions between the factors of $\slashed{a}(n.x)$ on either side of the photon line, see Fig.~\ref{FIG:WARD}. This generates an additional term at each vertex which is missed if one simply begins with the averaged propagator~(\ref{G2VSR}). Dropping the rapidly oscillating terms and going to momentum space, the structure of the three-point vertex becomes 
\be\label{GAMMA-1}
S_\text{vsr}(q+k).\Gamma^\mu (q+k,q).S_\text{vsr}(q)G_{\mu\nu}(k) \;, \quad \text{ where } \quad \Gamma^\mu (q+k,q) = \gamma^\mu +\frac{\delta m^2 n^\mu}{2n.(q+k) n.q} \slashed{n} \;.
\ee
$\Gamma$ is precisely the three-point vertex of SIM(2)--QED, and the second term in $\Gamma^\mu$ arises from the contractions of $\slashed{a}(n.x)$. Further, if we amputate the photon propagator and replace it with $k_\mu$ then we find by direct computation that
\be
	\Gamma^\mu(q+k,q) k_\mu = {S}_\text{vsr}(q) - {S}_\text{vsr}(q+k) \;,
\ee
which is the Ward-Takahashi identity; without the second term in (\ref{GAMMA-1}) this would be violated. Hence the averaging approximation preserves gauge invariance provided we are careful to retain all the slowly varying terms. In doing so each QED vertex becomes the three-point vertex of SIM(2)--QED.

\subsection{$n$-point vertices}
In SIM(2)--QED in a general covariant gauge there are an infinite number of vertices at which two fermion lines meet with arbitrary numbers of photon lines: showing how these vertices arise is the final step in establishing the correspondence between our theories. Consider then the short-distance behaviour of the background field fermion propagator~(\ref{GVOLK}).  $S$~contains a term which is local in~$n.x$, referred to as the `instantaneous electron propagator' in lightfront field theory~\cite{Brodsky:review,Heinzl:review}. This term hides a slowly varying contribution which may be isolated by temporarily dividing up the background field propagator using the identity
\be\label{uppdelningen}
	\frac{\slashed{p} + m}{p^2-m^2+i\epsilon} = \frac{\slashed{p}_\text{o.s.} + m}{p^2-m^2+i\epsilon} + \frac{\slashed{n}}{2n.p} \;,
\ee 
in which
\be
	p_\text{o.s.}^\mu := p^\mu - \frac{p^2-m^2}{2n.p} n^\mu \;,
\ee
is on-shell. Inserting the decomposition (\ref{uppdelningen}) into (\ref{GVOLK}) and performing the $\bar{n}.p$ integral (where $\bar{n}^2=0$ and $\bar{n}.n=0$) the first term of (\ref{uppdelningen}) gives a propagator proportional to $\theta(n.x-n.y)$~\cite{Brodsky:review} and therefore does not generate any slowly varying terms beyond those already discussed. The second term of (\ref{uppdelningen}) gives the instantaneous propagator $I(x-y)$,
\be\label{S-INST}
	I(x-y) = i\int\!\frac{\ud^4p}{(2\pi)^4} \frac{\slashed{n}}{2n.p} e^{-ip.(x-y)} \propto \delta(n.x-n.y) \;.
\ee
In a correlation function with an internal fermion line between two vertices at $x^\mu$ and $y^\mu$ the instantaneous propagator appears as 
\be
\label{INST-BIDRAG0}
	\int\ldots \phi^\dagger_{p_\text{out}}(x)\bigg(1+\frac{\slashed{a}(n.x)\slashed{n}}{2n.p_\text{out}}\bigg) \gamma^\mu I(x-y) \gamma^\nu \bigg(1+\frac{\slashed{n}\slashed{a}(n.y)}{2n.p_\text{in}}\bigg)\phi_{p_\text{in}}(y) \ldots
\ee
There are again two slowly varying terms, the first coming from the factors of (matrix) unity, the second coming from the product of the terms linear in $\slashed{a}$: writing $l_\mu$ for the propagator momentum, the spinor structure is
\be
\label{INST-BIDRAG}
	(\ref{INST-BIDRAG0}) \sim \int\ldots \delta(n.x-n.y)\frac{1}{n.l}\bigg(\gamma^\mu \slashed{n}\gamma^\nu  + \frac{\delta m^2 n^\mu n^\nu}{2n.p_\text{out} n.p_\text{in}}  \slashed{n} \bigg) + \ldots \;.
\ee
up to rapidly oscillating terms. In the large brackets, the first term is a free-theory contribution, independent of $a_\mu$. It is unaffected by averaging, and can be recombined with the `on-shell' part of the propagator in (\ref{uppdelningen}), with the result that the sum of their contributions reproduces the SIM(2) fermion propagator (\ref{G2VSR}) between two three-point vertices. In other words, (\ref{G2VSR}) is still the propagator which appears between the three-point vertices in all diagrams, and the discussion above of contractions across photon lines still holds.
\begin{figure}[t!]
	\centering\includegraphics[width=0.4\columnwidth]{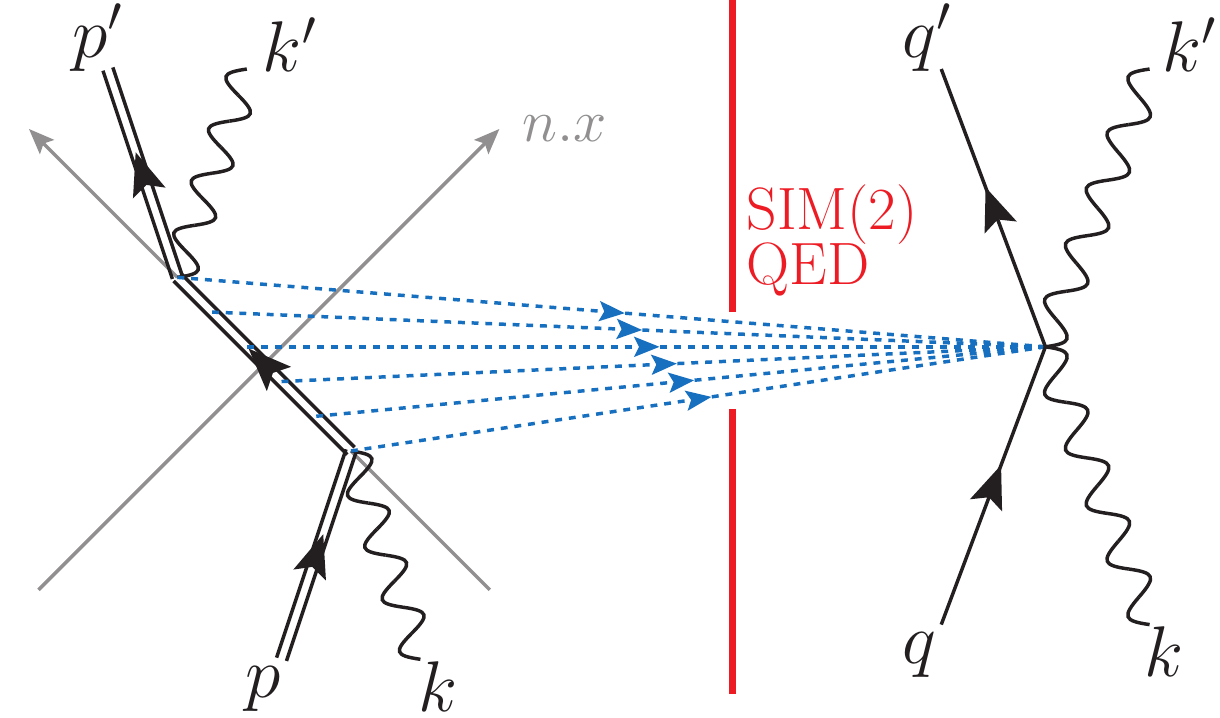}
	\caption{\label{FIG:INST} Propagation of the background-field dressed fermion at constant $n.x$ in QED, left, gives slowly varying terms which generate the higher-order vertices of SIM(2)--QED, right.}
\end{figure}

The second term in the large brackets of (\ref{INST-BIDRAG}) is the `hidden' $a^2$ contribution from the instantaneous propagator which is not explicit in (\ref{GVOLK}). One such term appears per internal line, and products of such terms give the higher-order vertices of SIM(2)--QED: for a single line, the delta function in $n.x$ brings together two vertices and we obtain a (2-fermion) 2-photon vertex, for two connected internal lines the product of two delta-functions pulls three vertices together, giving a 3-photon vertex, and so on. We will demonstrate this explicitly with Compton scattering, which illustrates all the important points and allows a direct comparison with the calculations in~\cite{Dunn}.
\subsection{Example: Compton scattering}
Consider then Compton scattering in QED, within the background wave, as shown in Fig.~\ref{FIG:INST}:
\be
	e(p,s) + \gamma(k,\epsilon) \overset{\text{in background}}{\longrightarrow} e(p',s') + \gamma(k',\epsilon') \;. 
\ee
(We suppress spin labels from here on.) The $S$-matrix element is
\be
\begin{split}
	S_{fi} = (-ie)^2 \int\!\ud^4x\ud^4y\;  &\bar{u}_{p'}\bigg(1 -\frac{\slashed{n}\slashed{a}(n.x)}{2n.p'}\bigg)\phi_{p'}^\dagger(x) \slashed{\epsilon}'e^{ik'.x} S(x,y) \slashed{\epsilon} e^{-ik.y} \phi_p(y) \bigg(1 + \frac{\slashed{n}\slashed{a}(n.y)}{2n.p}\bigg) u_p 
	\\ &+ (\epsilon' \leftrightarrow \epsilon , k \leftrightarrow -k' \big) \;,
	\end{split}
\ee
where the final term represents the photon exchange diagram\footnote{The calculation of probabilities and cross-sections from such background field amplitudes with higher numbers of vertices can be challenging, especially when one considers nontrivial background field profiles. See~\cite{Hartin:thesis,Hu:trident,Ilderton:trident,Seipt:dubbel,Mackenroth:dubbel,King:trident} for recent examples and discussions, as well as references therein.}. We proceed to identify all the slowly varying terms. Using the separation (\ref{uppdelningen}) splits the propagator into two terms; writing $l_\mu$ for the propagator momentum again, the slowly varying contributions from the `on-shell' term are
\be\label{trams1}
	\begin{split}
		S_{fi} \supset i(-ie)^2& \int\!\ud^4x \ud^4y \int\!\frac{\ud^4 l}{(2\pi)^4}\; e^{i(q'+k'-l).x -i(q+k-l).y + \ldots} \times \\
		&\times\bar{u}_{p'} \epsilon'_\mu  \bigg(\gamma^\mu +\frac{\delta m^2 n^\mu}{2n.p' n.l} \slashed{n} \bigg) \frac{\slashed{l}_\text{o.s.}+m}{l^2-m^2+i\epsilon}
	 \bigg(\gamma^\nu +\frac{\delta m^2 n^\nu}{2n.l n.p} \slashed{n} \bigg) \epsilon_\nu  u_p + \ldots 
	\end{split}
\ee
where the ellipses denote rapidly varying terms, as well as the exchange terms. Discarding the rapidly varying terms leaves us with two SIM(2) three-point vertices connected by a propagator which contains $\slashed{l}_\text{o.s.}$ rather than the off-shell $\slashed{l}$: this is not the SIM(2) propagator. The latter will reemerge, though, once we add the contribution from the instantaneous propagator (\ref{S-INST}):
\be\label{trams2}
\begin{split}
	S_{fi} \supset i(-ie)^2& \int\!\ud^4x\ud^4y\int\!\frac{\ud^4l}{(2\pi)^2}\,  e^{i(q'+k'-l).x -i(q+k-l).y + \ldots} \times \\
	&\bar{u}_{p'} \bigg( \slashed{\epsilon}' \slashed{n} \slashed{\epsilon}  - \frac{\slashed{a}(n.x) \slashed{a}(n.y) n.\epsilon' n.\epsilon}{2n.p' n.l n.p}\slashed{n}\bigg) u_p + \ldots
\end{split}
\ee
Summing the first term with (\ref{trams1}) replaces $\slashed{l}_\text{o.s.} \to \slashed{l}$ in the averaged result, recovering the SIM(2) propagator connecting two SIM(2) three-point vertices:
\be\label{trams3}
	\begin{split}
		S_{fi} \supset -ie^2& (2\pi)^4\delta^4(q+k'-q-k)\bar{u}_{p'} \epsilon'.\Gamma(q',q+k) \tilde{S}_\text{vsr}(q+k)
	 \epsilon.\Gamma(q+k,q) u_p \\
	 &+ (\epsilon' \leftrightarrow \epsilon , k \leftrightarrow -k' \big)
	\end{split}
\ee
Turning to the second term in (\ref{trams2}), we perform the integrals to find, including the exchange term,
\be\label{COMPTON-FYRA}
	S_{fi} \supset -ie^2(2\pi)^4 \delta^4(q'+k'-q-k) \frac{\delta m^2}{2}\frac{n.\epsilon' n.\epsilon}{n.q'n.q}\bigg(\frac{1}{n.(q+k)} + \frac{1}{n.(q-k')}\bigg)\bar{u}_{p'} \slashed{n} u_p \;.
\ee
The sum of (\ref{trams3}) and (\ref{COMPTON-FYRA}) is the Compton scattering amplitude in SIM(2)--QED, and from (\ref{COMPTON-FYRA}) we read off an effective four-point vertex which may, using the momentum-conserving delta-function, be written
\be\label{FYRA}
	ie^2\frac{\delta m^2 n_\mu n_\nu \slashed{n}}{2 n.k\, n.k'} \bigg( \frac{1}{n.q} + \frac{1}{n.q'} - \frac{1}{n.(q+k)} - \frac{1}{n.(q-k')}\bigg) \;.
\ee
This is exactly the 2-photon 2-fermion vertex in SIM(2)--QED, see e.g.~\cite{Dunn} (the notation of which is recovered by making the replacements, in (\ref{FYRA}), $\{q,q',k,k'\} \to \{ p,p',q_1,-q_2\}$). Similarly, diagrams with two or more internal fermion lines generate, via the products of instantaneous terms in multiple propagators, the local SIM(2)--QED vertices at which two or more photon lines meet.

We have thus shown that the {\it effective} description of QED in a very high frequency background wave is SIM(2)-QED in VSR. The limits of this correspondence are implicit in the phrase `high frequency'. As the background is a toy model for some unknown \ae ther, $\omega$ introduced above should be set at some beyond-the-Standard-Model scale such that, in the lab frame where the background has the form (\ref{bakgrund}), $\omega$ is much larger than achievable particle energies. Then it makes sense to average out the oscillations, as we have done. When particle energies reach the same scale as~$\omega$, though, they will probe the rapid oscillations of the background and one will clearly need more than just the effective theory to capture all the physics. Classically, this is illustrated by the above comparison of the physical and effective masses, see (\ref{massskiften}), above.

\subsection{Lightfront gauge}\label{SECT:LF}
Like the $\delta m^2$-dependent terms in the three-point vertex, the presence of the higher-order vertices is essential for preserving gauge invariance in SIM(2)--QED: for example the Compton amplitude, write it $\epsilon.\mathcal{M}$, only obeys the Ward identity $k.\mathcal{M}=0$ if both the 1-photon and 2-photon vertices are included. Suppose though that we had performed only the {\it naive} averaging of the propagator~(\ref{G2VSR}) and neglected the contributions (\ref{GAMMA-1}) and (\ref{INST-BIDRAG}) shown in Fig.'s~\ref{FIG:WARD} and~\ref{FIG:INST}. The Ward identities would then have been violated, and one way to preserve them would have been to add new vertices to the theory -- in this way the higher-order vertices could have re-emerged. In other words, had we only modified the Dirac equation, we would have been forced to also modify Maxwell's equations: compare~\cite{Alfaro:BMT} where a modified supersymmetry constraint for the VSR spinning particle necessitates modified Maxwell's equations.

Having now seen how the vertices arise and how gauge invariance is preserved, we can consider the theory in different gauges.  Observe that all the higher order vertices will be, as in (\ref{COMPTON-FYRA}), proportional to $n^\mu$, and hence will drop out of all correlation functions if we go to lightfront gauge, $n^\mu G_{\mu\nu}(k)=0 = n_\mu \epsilon^\mu(k)$~\cite{Dunn,Cheon,S2}. We then recover the Feynman rules of SIM(2)--QED in lightfront gauge: the fermion propagator is (\ref{G2VSR}), the three-point vertex reduces to the usual $-ie\gamma^\mu$, and all the higher-order vertices disappear, being in effect shuffled into the photon propagator which itself acquires an instantaneous part~\cite{Brodsky:review},
\be\label{FR-VSR-2}
	G_{\mu\nu}(k) = \frac{-i}{k^2+i\epsilon} \bigg( g_{\mu\nu} - \frac{n_\mu k_\nu + k_\mu n_\nu}{n.k} \bigg) \;.
\ee
The propagators and vertices above contain $n.p$ terms in denominators. Because propagator momenta are off-shell these terms can (unlike in the case of classical motion in Section~\ref{SEKT:DIRAC}) go to zero and cause divergences.  These `zero mode' singularities are well known from lightfront field theory, reviewed in e.g.~\cite{Brodsky:review,Heinzl:review}, where they have a long history and are intrinsically linked to the structure of the vacuum~\cite{Heinzl:1991vd}. They are commonly regularised using a principal value prescription, $1/n.p \to \mathcal{P}(1/n.p)$ (and this has been taken up in VSR~\cite{Dunn}) though a variety of related but more sophisticated prescriptions offer good properties such as preservation of causality~\cite{Mandelstam:1982cb,PhysRevD.29.1699,PhysRevD.42.2115}. The zero modes are a subtle issue even in QED, and in general they cannot simply be excluded~\cite{Heinzl:1991vd,Ji:1995ft}: effects such as nonperturbative pair production are driven by zero mode contributions, and this physics is lost if the zero modes are removed~\cite{Tomaras:2001vs,Ilderton:2014mla}. For our purposes, though, we need only note that the chosen regularisation in QED is inherited by SIM(2). 

The relevance of lightfront methods is further seen by recalling that the stability group of the lightfront quantisation surface, $n.x=0$, is the largest in all possible quantisation schemes, being 7-dimensional~\cite{Dirac:dynamics,Leutwyler:dynamics} and is generated precisely by SIM(2) together with one longitudinal and two transverse translations. Lightfront methods have also proven useful in establishing holographic dualities, as reviewed in~\cite{BrodskyQCD}. Recalling that the background field amplitudes above are relevant to laser-matter interactions~\cite{Marklund:2006my,Heinzl:ELI,DiPiazza:review}, the particular role of lightfront methods has been highlighted in~\cite{Neville,Bakker}.

\section{Conclusions}\label{SEKT:CONCS}
QED in a high-frequency background, averaged over rapid field oscillations, is SIM(2)-invariant, and $\mathcal C$-invariant, QED. Thus Very Special Relativity can emerge as an effective description of a background field theory below the high frequency scale of the background. That the Lorentz-violating terms in SIM(2) can be nonlocal~\cite{original,Dunn} is explained in this approach by the fact that they survive an averaging, which is a nonlocal operation. Averaging also restores translation invariance to the effective theory, so that momentum becomes a good quantum number, in agreement with VSR. The higher-order vertices of SIM(2)--QED are correctly generated and the Ward identities are preserved.

There are many possible topics for future study. We have considered only a single field model, but the approach taken here should extend to other field profiles provided the background remains very rapidly oscillating compared to typical frequency/energy scales. One could consider the phenomenological impact of allowing for a slowly varying pulse envelope on top of the rapid oscillations (here we have considered a flat-top envelope), so that $\delta m^2$ would become a function of the lightlike direction. Effective masses as in (\ref{massskiften}) can appear for other field profiles, for example, and the effective momenta (\ref{q}) can acquire a richer structure~\cite{Harvey:massa}; it would be interesting to see if the resulting effective theories also have a VSR-like interpretation. 

Other possible topics include the extension of the ideas presented here to other parts of the SIM(2)-invariant standard model~\cite{VSR-SM}, Born-Infeld electrodynamics~\cite{VSR-BI}, noncommutativity~\cite{NC} and other VSR groups~\cite{original,S1}. One of the intriguing properties of VSR theories is that they provide a mechanism for neutrino mass-generation~\cite{Cohen2,Dunn}. Neutral particle masses clearly cannot arise at tree level using the present approach, as the amplitude $\delta m$ which parameterises Lorentz-invariance-violating effects is proportional to the particle charge. It would though be interesting to investigate whether neutrino mass terms could arise radiatively, through the coupling of neutrinos to charged particles which in turn couple to the electromagnetic background. \\

\acknowledgements
\textit{A.I.~thanks Tom Heinzl and Greger Torgrimsson for useful discussions. A.I.~is supported by the Olle-Engkvist Foundation, grant 2014/744.}

\bibliography{VSR-bib}
\end{document}